\documentclass[12pt]{article}%

\usepackage{subcaption}
\usepackage{amssymb}
\usepackage{hhline}
\RequirePackage[OT1]{fontenc}
\RequirePackage{amsthm,amsmath}
\usepackage[authoryear]{natbib}
\RequirePackage[colorlinks,citecolor=blue,urlcolor=blue]{hyperref}
\usepackage[margin=1in]{geometry}
\usepackage{graphicx}
\usepackage{authblk}
\allowdisplaybreaks
\usepackage{float}
\usepackage{setspace}
\usepackage{multirow}
\usepackage{algorithm}
\usepackage{algpseudocode}

\theoremstyle{plain}

\restylefloat{table}

\newcommand{\btheta}{ \mbox{\boldmath $ \theta $} }

\newcommand{\bone}{\textbf{1}}

\newcommand{\bE}{\textbf{E}}

\newcommand{\bw}{\textbf{w}}

\begin{document}

\title{Generalized Evolutionary Point Processes: Model Specifications and Model Comparison}
\author[1]{Philip A. White \thanks{Corresponding Author: pwhite@stat.byu.edu}}
\author[2]{Alan E. Gelfand \thanks{alan@duke.edu}}

\affil[1]{Department of Statistics, Brigham Young University, Provo, UT, USA}
\affil[2]{Department of Statistical Science, Duke University, Durham, NC, USA}

\maketitle

\begin{abstract}

Generalized evolutionary point processes offer a class of point process models that allows for either excitation or inhibition based upon the history of the process. In this regard, we propose modeling which comprises  generalization of the nonlinear Hawkes process. Working within a Bayesian framework, model fitting is implemented through Markov chain Monte Carlo.  This entails discussion of computation of the likelihood for such point patterns.   Furthermore, for this class of models, we discuss strategies for model comparison. Using simulation, we illustrate how well we can distinguish these models from point pattern specifications with conditionally independent event times, e.g., Poisson processes.  Specifically, we demonstrate that these models can correctly identify true relationships (i.e., excitation or inhibition/control). Then, we consider a novel extension of the log Gaussian Cox process that incorporates evolutionary behavior and illustrate that our model comparison approach prefers the evolutionary log Gaussian Cox process compared to simpler models.  We also examine a real dataset consisting of violent crime events from the 11th police district in Chicago from the year 2018. This data exhibits strong daily seasonality and changes across the year. After we account for these data attributes, we find significant but mild self-excitation, implying that event occurrence increases the intensity of future events.

\end{abstract}

\noindent\textsc{Keywords}: {

Bayesian framework, conditional intensity, inhibition, Markov chain Monte Carlo, (nonlinear) Hawkes process, self-exciting

}





\section{Introduction}\label{sec:intro}

Self-exciting point process models, in particular the Hawkes process, are a class of point processes that capture excitatory relationships among observed events  \citep{hawkes1971b,hawkes1971a}. More precisely, the occurrence of an event in the window $(t, t + dt]$ is encouraged by the occurrence of events recent to time $t$.  The Hawkes process has come to prominence for modeling earthquakes \citep[see][]{ogata1988,ogata1998,fox2016}, crime events \citep[see, e.g.,][]{mohler2011}, financial trading \citep[see][]{embrechts2011}, and neurophysiology \citep[see][for an early discussion]{ chornoboy1988}. More recently, these models have been adopted and extended in the machine learning literature to quantify the interactions between events on network structures \citep[see, e.g.,][]{ blundell2012, farajtabar2017,chen2018b}.

In many settings, \emph{strict} self-excitation is too limiting. With a model that allows for both excitation and no excitation, \cite{linderman2014} argue for a spike-and-slab estimation approach for inferring about excitation parameters. However, the presence or absence of excitation does not allow for identifying the presence of inhibitory relationships between events.

Some authors have lifted the constraint of strict excitation but have failed to insist that the intensity be nonnegative \citep[see, e.g.,][]{pillow2008}. In neurophysiology, \cite{hansen2015,mei2017,chen2017} agree that strict excitation is too limiting because inhibition is an important relational component in neuronal networks.  In doing so, they offer a variety of coherent remedies. \cite{hansen2015} propose a simple \emph{rectifier function} $\max(x,0)$ around the intensity to force the intensity to be non-negative. We refer to this as a Tobit link \citep{tobin1958} specification. \cite{chen2017} propose constrained versions of the Hawkes process that allow for inhibitory relationships in a theoretically coherent way. However, their approach requires a complicated iterative construction procedure. \cite{mei2017} argue for an intuitive approach that uses a scaled softplus function, e.g., $\text{ln}(1 + e^{x})$, as a link function for an unconstrained intensity, in order to restrict the intensity to be positive. We generalize the approach of \cite{mei2017} to consider more general link functions.

Our contributions are as follows.  First, we propose straightforward and intuitive evolutionary point process models without aggregating to arbitrary time windows. For this, we generalize nonlinear Hawkes processes to also include dampening relationships. Next, in this context, we discuss model fitting in a fully Bayesian framework that allows for full posterior inference regarding parameters along with associated uncertainty. Further, we propose model comparison criteria appropriate for evolutionary point processes.  Then, we employ a simulation study to illuminate when and how well we can identify departure from conditionally independent event times in favor of excitation or inhibition. As a part of this, we introduce a novel extension of the log Gaussian Cox process that includes an evolutionary component. Lastly, we consider a real dataset and, using this evolutionary log Gaussian Cox process, reveal the presence of mild self-excitation with regard to event times.

We define an \emph{evolutionary point process} as a point process model whose specification relies on previously observed points, i.e., is specified through a \emph{conditional intensity}; we borrow this terminology from \cite{rasmussen2011}.  Within this modeling framework, we propose useful metrics for model comparison. Previous approaches for comparing evolutionary point process models, in particular, Hawkes process models, include using the Akaike information criterion (AIC) \citep{ogata1988} or employing simulations from the model \citep{chen2018b} using mean squared error. Strategies we eschew are the use of the random time change theorem \citep[see, e.g.,][]{meyer1971,ogata1988} and Pearson-like residuals \citep[see, e.g.,][]{bray2013}.  We avoid these approaches because they first require estimation of the compensator or the intensity before comparing the data to model predictions.  Hence, they use the data twice, weakening model criticism.   We also note the hypothesis testing approach of \cite{dachian2008}.  They consider formal locally most powerful testing for departure from a Poisson process in the direction of self-exciting behavior.  We work within a Bayesian framework, with explicit classes of parametric models, and with model comparison in the data space so our perspective is not directly comparable with theirs.


The format of the paper is as follows. We introduce evolutionary modeling in the context of current modeling approaches in Section \ref{sec:point_patterns}. We discuss generalization of the nonlinear Hawkes process that permits both excitatory and inhibitory relationships in Section \ref{sec:model}.  
In response to unique challenges associated with evolutionary point processes, we propose model comparison approaches in Section \ref{sec:mod_comp_val}. In Section \ref{sec:sim}, we provide simulation studies that explore what aspects of the model we can effectively discriminate using model comparison. Then, in Section \ref{sec:data}, we provide an analysis of violent crime events in the 11th Police District in Chicago, Illinois, USA in the year 2018  where we find mild self-excitation of event times, even accounting for other data characteristics. We conclude with a summary and possible extensions of our modeling framework in Section \ref{sec:conc}.

\section{Evolutionary Point Processes }\label{sec:point_patterns}


For a realization of a point process, we denote $N(\cdot)$ as the counting measure over $(0,T]$, where, for any Borel set $A \subset (0,T]$, $N(A) = \sum_{t} \bone(t_i  \in A)$ and $\bone$ is the indicator function.  We define an evolutionary point process through a conditional intensity, that is, its intensity at time $t$ depends upon the history of the process up to time $t$. Let $\mathcal{T} = \{t_1,...,t_n \}$ to be an observed point pattern of events, where $t_i \in (0,T] \subset \mathbb{R}$ is an event time. We let the history $\mathcal{H}(t)$ of the point pattern up to time $t$ be $\mathcal{H}(t) \equiv \{ t_k : t_k < t \}$.

The conditional intensity, given the history up to time $t$, is denoted by $\lambda^*(t) = \lambda(t | \mathcal{H}(t))$ and captures the (instantaneous) expected number of points per unit time. These types of point processes allow the intensity to depend directly upon event occurrence, enabling either excitation or dampening of the intensity. Here, we provide a few examples from each class but do not attempt an exhaustive review.

The \emph{compensator} asssociated with a point process is customarily defined as $\Lambda(t) = \int_{0}^{t} \lambda(s) ds$.  Applied to an evolutionary point process, we denote it by $\Lambda^*(\cdot)$ with
\begin{equation}
\Lambda^*(t)  = \int^t_0 \lambda^*(s) ds = \int_{0}^{t} \lambda(s | \mathcal{H}(s))ds.
\end{equation}
The compensator is a cumulative intensity and gives the expected number of points up to time $t$.
For most point process models, the intensity has unknown parameters, which we denote as $\btheta$ and, accordingly, we write $\lambda(\cdot | \btheta)$, or, for a conditional intensity, $\lambda^*(\cdot | \btheta)$.

Turning to self-exciting point process models, the Hawkes process \citep{hawkes1971a} is the most common. For a point pattern $\mathcal{T}$ with events $t_i \in \mathcal{T}$, the conditional intensity of the Hawkes process is written as
\begin{equation}\label{eq:hawkes_3}
    \lambda^*(t; \btheta) = \mu(t; \btheta_{1}) + \sum_{t_i < t} g(t-t_i; \btheta_{2}) ,
\end{equation}
where $\mu(t; \btheta_{1}) > 0 $ is a parametric specification for the background intensity and $g(\cdot; \btheta_{2}): \mathbb{R} \mapsto \mathbb{R}^{+}$ is the parametric excitation intensity.  So, altogether,  $\btheta = (\btheta_{1}, \btheta_{2})$.

In most work the background intensity $\mu(t)$ is assumed to be constant, ignoring potentially relevant covariate information. \cite{simma2012} propose using using covariate information in both $\mu(t)$ and $g(\cdot)$. This is a potentially important modeling aspect but it is beyond the scope of this manuscript.

Some possible choices of $g(\cdot; \btheta_{2})$ are given in Table \ref{tab:hawkesg}. As was done by \cite{rasmussen2013}, we decompose the excitation (or triggering or offspring) function into a density ($f(\cdot)$) times a multiplicative factor ($\alpha >0$). That is, $g(\cdot) =  \alpha f(\cdot)$. In the examples given in Table \ref{tab:hawkesg}, $\alpha$ serves as a scaling factor to the normalized offspring intensity $f(\cdot)$. Though some applications might encourage alternative choices for $g$, in the sequel we confine ourselves to the exponential.

\begin{table}[H]

\centering

\begin{tabular}{llr}

  \hline

 & Function $g(t)$ & Parameters  \\

  \hline

Exponential & $\alpha \beta e^{-\beta t}$ & $\alpha > 0$, $\beta >0$  \\

Uniform & $\alpha / \beta $ & $\alpha > 0$, $\beta >0$, $\bone(t \in (0,\beta) )$ \\

Gaussian & $\alpha  (2 \pi \beta^{-1})^{-1/2} e^{- \frac{\beta t^2}{2}}$  & $\alpha > 0$, $\beta >0$ \\

Triangle & $\frac{2\alpha}{\beta} \left(1 - \frac{t}{\beta}\right)$ & $\alpha > 0$, $\beta >0$, $\bone(t \in (0,\beta) )$ \\

Gamma & $ \alpha \frac{\beta^\nu}{\Gamma(\nu)} t^{\nu -1} e^{-\beta t}$ & $\nu >0$, $\alpha > 0$, $\beta >0$ \\

   \hline

\end{tabular}

\caption{Possible triggering functions $g(t)$, where $t$ represents the time difference between two points. }\label{tab:hawkesg}

\end{table}

The assumption of additive excitation, as in \eqref{eq:hawkes_3} may not be desirable or justified by the application. In this case, the nonlinear Hawkes process can be used \citep[see][for fuller discussion on nonlinear Hawkes processes]{bremaud1996,zhu2013}. For a monotonic function $h: \mathbb{R}^{+} \mapsto \mathbb{R}^{+}$, the nonlinear Hawkes process has a conditional intensity function,
\begin{equation}\label{eq:nonlinear_hawkes}
\lambda^*(t) =  h\left( \mu(t; \btheta_{1})  +  \sum_{t_i < t} g(t - t_i ; \btheta_{2}) \right),
\end{equation}
where the components of the intensity are analogous to those in (\ref{eq:hawkes_3}), usually with the same constraints. This is still a strictly self-exciting point process model; $h(\cdot)$ merely modulates the conditional intensity across $t$. Choices for $h(\cdot)$ are considered in the next section.

Self-controlling (or self-regulating) point processes \citep{isham1979} can be viewed as an extension of the nonlinear Hawkes processes in \eqref{eq:nonlinear_hawkes}. These processes, unlike self-exciting point processes, are applied in settings where the occurrence of events up to time $t$ are assumed to decrease the incidence of occurrence of events in the interval $(t, t + dt]$. A model of this type has been used, for instance, in modeling queues \citep{daley2003}. An example of a suitable intensity is
\begin{equation}\label{eq:self_control}
\lambda^*(t) = \exp\left( \mu t  -  \beta N((0,t)) \right),
\end{equation}
where $\mu, \beta > 0$, and $N(\cdot)$ is the counting measure. In this case, $\mu$ and $\beta$ compete, $\mu$ increasing and $\beta$ decreasing the intensity. Ostensibly, this model could be adapted to capture opposing phenomena, \emph{self-activation}, where $\lambda^*(t)$ decreases with $t$ until events excite the intensity.


The likelihood for realization $\mathcal{T}$ under a general conditional intensity, with associated compensator, becomes
\begin{equation} \label{eq:like}
e^{-\Lambda^*(T)} \prod_i \lambda^*(t_i),
\end{equation}
which is then used for inference about an evolutionary point pattern.  Derivation of this likelihood is provided in Appendix \ref{app:likelihood}.  






\section{General Evolutionary Point Process Models}\label{sec:model}

\begin{figure}[H]
 \begin{center}
     \includegraphics[width=0.48\textwidth]{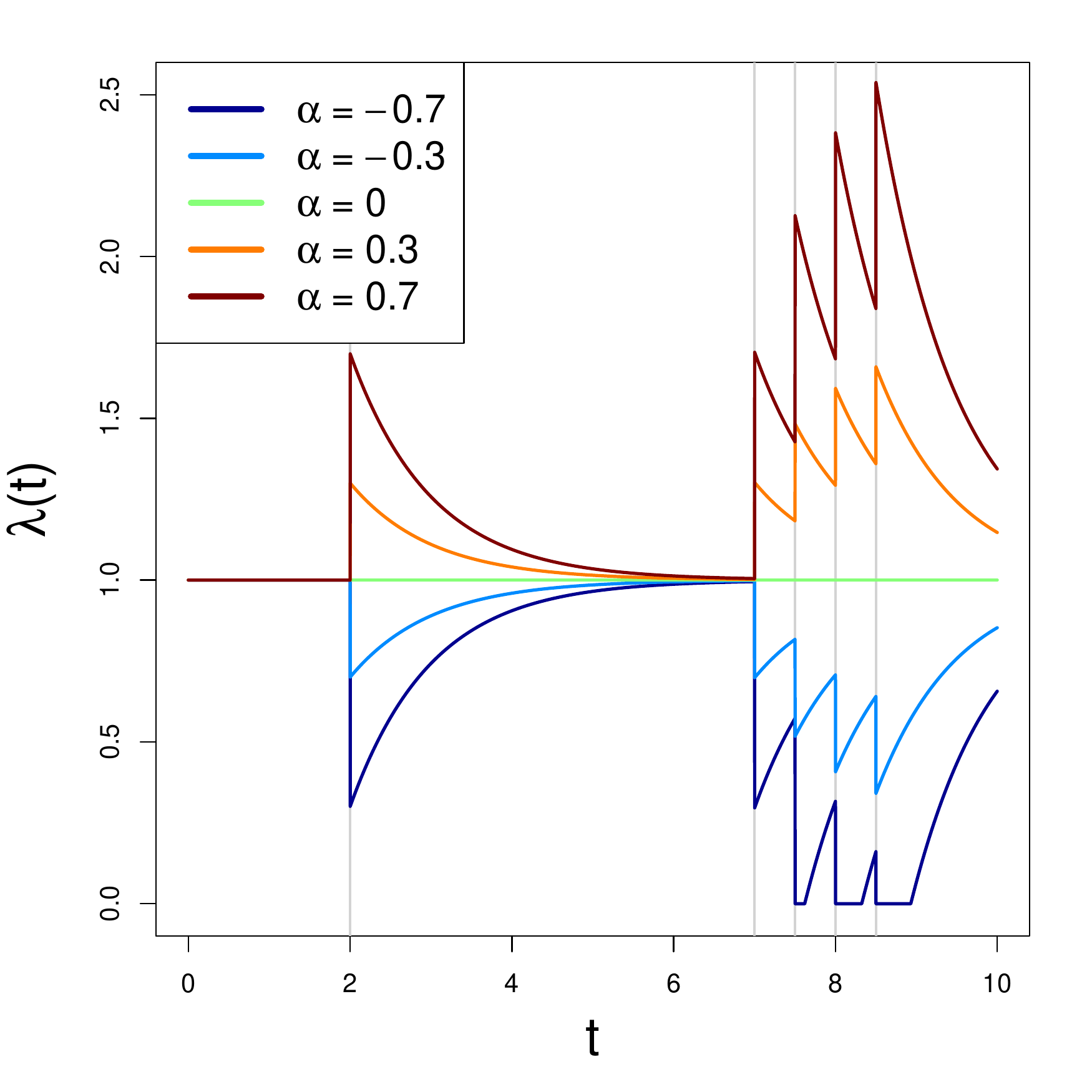}
          \includegraphics[width=0.48\textwidth]{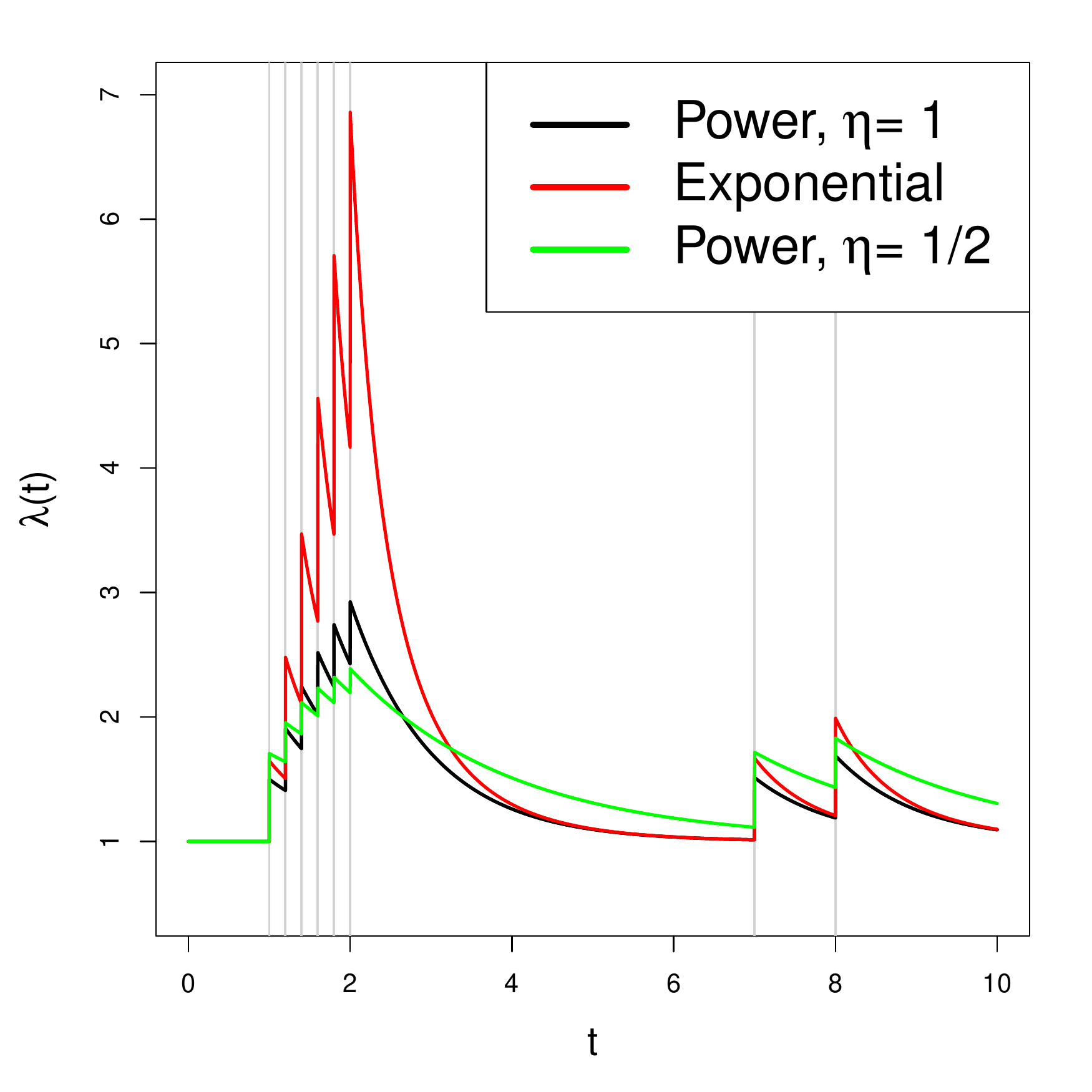}

\caption{ (Left) Self-exciting intensities with different $\alpha$s for a Tobit $h(\cdot)$. Events, indicated by a gray line, are at times 2, 7, 7.5, 8, 8.5, 9. (Right) Self-exciting intensities with different $h(\cdot)$. Events, indicated by a gray line, are at times 1, 1.2, 1.4, 1.6, 1.8, 2, 7, 8).}\label{fig:spectrum}
    \end{center}
\end{figure}

To specify general evolutionary point process models, we adopt versions of the nonlinear Hawkes process \eqref{eq:nonlinear_hawkes}, allowing the triggering function to be both positive and negative and also using a monotonic function $h(\cdot)$ that maps the unconstrained intensity to $\mathbb{R}^{+}$. That is, for $h: \mathbb{R} \mapsto \mathbb{R}_+$, the conditional intensity becomes
\begin{equation}\label{eq:general_univariate}
\lambda^*(t) =  h\left( \mu(t; \btheta_{1})  +  \sum_{t_i < t} g(t - t_i; \btheta_{2} ) \right).
\end{equation}
Selections for $g(\cdot)$ presented in Table \ref{tab:hawkesg} are augmented so that we can now allow $\alpha \in \mathbb{R}$. To illustrate the flexibility of this model class, Figure \ref{fig:spectrum} shows realizations of $\lambda^{*}(t)$ for several different choices which are detailed in discussion below.


As discussed in Section \ref{sec:intro}, models of this type have been proposed (viz., $\max(x,0)$ or the scaled softplus function); however, other selections of $h(\cdot)$ may also be appropriate. We provide several examples of $h(\cdot)$ in Table \ref{tab:h_example}. Within this framework, modeling decisions focus on selection of $h(\cdot)$ and $g(\cdot)$.  The key issue concerns the interplay between $h$ and $g$. Can we distinguish $h$'s for a given $g$?  Can we distinguish $g$'s for a given $h$? This is the intent of our simulation study in the next section.  It is anticipated that a large number of event times will be needed to distinguish models.  Moreover, perhaps the more important objective is to be able to distinguish a nonlinear Poisson process, i.e., a process with $g=0$ from an exciting or from an inhibiting process.  So, this becomes the emphasis of our simulation investigation.  In the absence of covariates, a nonlinear homogeneous Poisson process is just a reparametrized homogeneous Poisson process.

We offer a brief technical paragraph.  For a model with $\int_{\mathbb{R}_+} |g(s)| ds < 0$, \cite{bremaud1996} shows that non-linear Hawkes processes are \emph{stable} only if the link function $h(\cdot)$ is $\alpha$-Lipschitz. Here, a stable model is one that, when simulated from, will not generate arbitrarily large point patterns.  Neural spike trains are often modeled as a generalized linear model using $h(\cdot) = \exp(\cdot)$, where the model is strictly excitatory, giving an extension of the homogeneous or nonhomogeneous Poisson process. However, these models are not stable because $\exp(\cdot)$ is not $\alpha$-Lipschitz \citep[see][for a discussion about the instability of this model]{gerhard2017}. For some remedies for the instability of these models, see \cite{chen2019}; however, this is not our focus. We note that no such issues arise for inhibitory models using $h(\cdot) = \exp(\cdot)$.  When $\alpha$ gives self-excitation, the power link in Table \ref{tab:h_example} yields a stable point process if $\eta \leq 1$. When $\eta > 1$, the process is not $\alpha$-Lipschitz and thus not stable. The remaining choices in Table \ref{tab:h_example} are $\alpha$-Lipschitz.

\begin{table}[H]
\centering
\begin{tabular}{ll}
  \hline
 & Function   \\
  \hline
Identity & $h(\cdot) = \cdot$   \\
Power Tobit/rectifier & $\left[\max \left( 0,\cdot \right)\right]^\eta$  \\
Soft-plus & $\ln\left(1 + e^{(\cdot)} \right)$  \\
$\log_{10}$ Soft-plus  & $\log_{10}\left(1 + e^{2.3(\cdot)} \right)$ \\
Exponential & $e^{(\cdot)}$  \\
   \hline
\end{tabular}
\caption{Possible link functions $h(\cdot)$ for the generalized evolutionary point processes.}\label{tab:h_example}
\end{table}

To visualize the flexibility of the models presented here, we plot several examples of intensities that arise from \eqref{eq:general_univariate} by changing $\alpha$ and $h(\cdot)$. Returning to Figure \ref{fig:spectrum}, in the left panel we plot intensities with $\alpha \in \{ -0.7,-0.3,0.0,0.3,0.7 \}$ with the same set of observed events, where we have $\mu = 1$, $g(\cdot)$ as an exponential with $\beta = 1$, and $h(\cdot) =  \max \left( 0,\cdot \right)$.
In the right panel of Figure \ref{fig:spectrum} we plot the intensity for three point processes that differ in terms of their link function $h(\cdot)$, the exponential link and the power link with $\eta = 1$ and $\eta = 1/2$. The exponential link shows greater self-excitement than the power link with $\eta = 1$. Because of the shape of the square-root function, the power link with $\eta = 1/2$ shows the more excitation than the power link with $\eta = 1$ when $\sum_{t_i <t} g(t - t_i) < 1$ but less excitation when $\sum_{t_i <t} g(t - t_i) > 1$. Moreover, the duration of the excitation depends on $h(\cdot)$. While excitation is most extreme with $h(\cdot) = e^{(\cdot)}$, it also decays most rapidly. On the other hand, excitation decays more slowly for the power link with $\eta = 1/2$. These plots illustrate the flexibility of these models; however, as we discuss in Section \ref{sec:sim}, it is difficult to discriminate among these link functions from a single point pattern.

\subsection{Prior Distributions and Model Fitting}\label{sec:prior_mod_fore}

To evaluate the likelihood \eqref{eq:like}, we must compute $\Lambda^*(T)$.  This expression is not generally analytically available because of the function $h(\cdot)$. Therefore, the integral is estimated numerically. Unlike the Hawkes process, there is no representation of this process as a cluster process \citep[see, e.g.,][]{hawkes1974,rasmussen2013}. However, there are other computational methods that can expedite likelihood computation. For example, with the exponential excitation function, the likelihood can be written using a recursion \citep{ogata1978} that enables likelihood computations of order $\mathcal{O}(n)$.
\cite{mei2017} use a Monte Carlo approximation for $\Lambda^{*}(T) = \int^T_0 \lambda^*(s) ds$. However, since this integral is only one-dimensional, we suggest that a simple numerical integration over a fine grid is faster and a better approximation than the Monte Carlo integral. We use the trapezoidal rule for numerical integration; we found no additional benefit compared to using Simpson's rule which employs a polynomial approximation to the function.

In the Bayesian framework, to completely specify the model, we also need to choose prior distributions for the parameters in $\mu(t; \btheta_{1})$ and $g(\cdot; \btheta_{2})$. In the examples we explore here, we assume that $\mu >0$\footnote{With some selections of $h(\cdot)$ (e.g. $\exp(\cdot)$), it would be appropriate to allow $\mu \in \mathbb{R}$. }.  Therefore, we propose Gamma and Log-Normal prior distributions for $\mu$. Similarly, except for $\alpha$, the parameters of $g(\cdot: \btheta_{2})$ are strictly positive; therefore, Gamma or Log-Normal prior distributions are proposed there, as well. For $\alpha$, we suggest normal or uniform prior distributions. We could also imagine using a mixture prior distribution with a prior point mass at $\alpha=0$ and a distribution for the rest of the support of $\alpha$ to see how much support the data offers for a homogeneous Poisson process.  This is similar to what \cite{linderman2014} proposed for the Hawkes process.

Our model fitting yields samples from the posterior distribution $[\btheta | \mathcal{T}]$ which are used to create realizations of the intensity function over $(0,T]$.  In Section \ref{sec:mod_comp_val},  we discuss how these realizations are employed for model comparison.  The models that we have presented here are too general to supply an overall Gibbs sampler. However, we do propose a Markov chain Monte Carlo (MCMC) model-fitting approach using the Metropolis-Hastings algorithm. To select proposal or candidate distributions for the parameters in the model, we use an adaptive MCMC where we begin our model fitting using a Metropolis-within-Gibbs update.  Then, while still in the burn-in period, we change to adaptive multivariate Metropolis updates, following the strategy presented in \cite{haario1999}. We detail our model fitting approach in Appendix \ref{app:model_fitting}.



\section{Model Comparison}\label{sec:mod_comp_val}



There is little literature on approaches for comparing evolutionary point process models and what exists is specifically for Hawkes process models.  Proposals include using the Akaike information criterion (AIC) \citep{ogata1988} or using simulations from the model \citep{chen2018b} with mean squared error.  We elect to implement such comparison in the data space rather than in the parameter space so that we can compare what is predicted under a model with what was observed.  Furthermore, when possible, we would prefer to do this out-of-sample, that is as a cross-validation, in order to avoid using the same data to fit the models and then also to compare models.  This enables more critical model comparison.

Because there is some tradition for model comparison in parameter space, we also include the deviance information criterion (DIC) results in our model comparison. For our setting, with $m$ MCMC samples, $\theta^{(1)},\theta^{(2)},...,\theta^{(m)}$, we define
$\bar{\theta} =\frac{1}{m} \sum^m_{i=1} \theta^{(i)}$ and
$\bar{D} =  \frac{1}{m} - 2 \log\{p(\mathcal{T} | \bar{\theta} ) \}$.
The effective model size or number of parameters is computed as  $p_D = \bar{D} - D(\bar{\theta})$. Then, the DIC is computed as $\text{DIC} = D(\bar{\theta})+ 2p_D = 2\bar{D} - D(\bar{\theta})$.
We will see that DIC performs well when comparing evolutionary point process models but fails to successfully discriminate evolutionary point process models from non-evolutionary models in some cases.

Since the intent of a self-exciting or self-inhibiting model is to capture short-term response at a given time to the history up to that time, we propose two approaches that rely on using the modeled conditional intensity to make short-term forecasts. The first approach uses the conditional intensity to approximate the probability of an event in a very short time window and compares this to the observed binary outcome of whether or not a point was present in that short window. The second uses the conditional intensity to obtain the expected number of events in a somewhat longer interval and compares this to the observed count of events in that interval. For the first criterion, consider a small change in time $\Delta_t$.  Then, since $\text{Pr}(N((t, t + \Delta_t]) >1$ is negligible,
\begin{equation}
\text{Pr}( N((t,t + \Delta_t])=1 ) \approx E(N((t,t + \Delta_t]) \approx  \lambda^*(t + \Delta_t/2) \, \Delta_t.
\label{eq:binapprox}
\end{equation}
Assuming $\lambda^*(t+ \Delta_{t}/2) \,\Delta_t \leq 1$, we have a model-based approximation to the probability of an event in a small window, $(t, t+ \Delta_{t}]$.  These probabilities can be compared with observed binary outcomes over a collection of short windows.  Note that we evaluate $\lambda^{*}$ at $t+\Delta_{t}/2$, i.e., a midpoint-rule approximation to $E(N((t,t + \Delta_t])$, rather than at $t$ because $\lambda^{*}(t)$ is not continuous at an event time $t$.

This procedure can be carried out \emph{retrospectively} or \emph{prospectively}. Retrospectively, one would fit a model to the entire point pattern, and then the model would be assessed based on that single fit. Prospectively, the model would be fit many times, each time up to a selected choice of $t$.  Here, for computational convenience, we use the retrospective approach.

What needs to be specified are the choices of the $t$'s and the choices of the $\Delta_t$'s, with no clear ``best'' strategy. We propose an approach motivated by distinguishing an evolutionary point pattern from a homogeneous Poisson process (HPP) with constant intensity $\lambda$. An evolutionary point process model, whether excitatory or inhibitory, behaves differently from an HPP right after an event. For a self-exciting model, we expect the intensity to exceed $\lambda$ shortly following events. In contrast, the intensity of inhibitory models drops below $\lambda$ following events. 
So, model comparison will be best served by using the set of observed event times $\mathcal{T}$

For each $t \in {\cal T}$ we seek to create two binary event probabilities, a $p_{t}$ arising under an HPP model and a $q_{t}$ arising under an evolutionary process model. We use $p_t$ and $q_t$ as probabalistice forecasts rather than converting them into a binary prediction. Thus, we implement model comparison by comparing the performance of the two sets of probabilities using probabilistic misclassification rates described below. To create these probabilities we first randomly draw a $\Delta_t$ to associate with each $t$. Let $\hat{\lambda} = n/T$ be the maximum likelihood estimate for $\lambda$ under the HPP. Then, we draw $p_t = \hat{\lambda} \Delta_t \sim \text{Unif}(0,1)$\footnote{Other selections may be justified. For example, self-exciting point processes can lead to $\lambda^*(t + \Delta_t/2) \Delta_t > 1$, even though $p_t$ is less than one. To prevent information loss, one may choose to sample $p_t \sim  \text{Unif}(0,a)$ for $a < 1$ or $p_t \sim  \text{Beta}(b,b)$ for $b > 1$; however, we found little benefit to these selections.}, so that the probability of an event in $(t, t + \Delta_{t}]$ is $p_t$ if the point pattern comes from an HPP. We then fix $\Delta_t = p_t / \hat{\lambda}$. We define $q_t = \min(1, \lambda^*(t + \Delta_t/2) \Delta_t)$ to provide the probability of an event in $(t, t+ \Delta_{t}]$ if the point pattern comes from the evolutionary process model. These probabilities are associated with binary events $Y_t$ for all intervals $(t,t + \Delta_t]$ following $t \in \mathcal{T}$. Here, we define $Y_t = 1$ if there is one or more event in $(t, t + \Delta_t]$ and $Y_t = 0$, otherwise. 

Then, to summarize model performance over intervals following $\mathcal{T}$, we employ a probabilistic misclassification rate, which we abbreviate as PMR. The definition for this misclassification rate depends on whether we are comparing self-exciting models or self-inhibiting models to an HPP. The intuition for a self-exciting model is that it will elevate probabilities following events in ${\cal T}$ if self-excitation is operating. Thus, misclassification would mean $Y_t = 1$ when $1 - q_t$ is large. The intuition for the inhibitory models is the opposite, we expect smaller probabilities following ${\cal T}$. In this case, misclassification occurs when $Y_t = 0$ but $q_t$ is large.
As a result, we have the following definitions of PMR for excitatory and inhibitory models, respectively:
\begin{align}\label{eq:tjurs}
\text{PMR}^{(\text{excite})} &= \frac{1}{\sum_t Y_t}  \sum_{t \in \mathcal{T}} (1 - q_t) Y_t \\
\text{PMR}^{(\text{inhibit})} &= \frac{1}{\sum_t (1 - Y_t)}  \sum_{t \in \mathcal{T}} q_t (1 - Y_t) . \nonumber
\end{align}
When comparing a self-exciting or self-inhibiting model to an HPP, we compute the same PMR for the HPP, replacing $q_t$ with $p_t$.

An attractive feature of this criterion is that we need not generate any posterior predictive realizations from the process.  We only need to obtain the posterior mean probability of $\lambda^*(t + \Delta_{t}/2)$. A potential criticism, as noted above, is the arbitrariness in the selection of the $t$'s. However, the randomization of the $\Delta_t$'s and focusing our model comparison on behavior following events, i.e., using $t \in \mathcal{T}$, helps to overcome this concern.  As a last remark here, misclassification rates are customarily applied to a specified set of events with an associated set of outcomes for these events. Here, we have to create our own suitable events.

Turning to the second approach, we would now use the approximation in \eqref{eq:binapprox} to estimate the expected number of points in the time window $(t, t + \Delta_t]$, rather than to create a probability. This provides a predictive distribution which is then compared to the observed number of events in the window. Again, there is the issue of choice of $t$ and $\Delta_t$.  Again, we use the event times in $\mathcal{T}$, as discussed above; however, we fix $\Delta_t = dt$ for all $t$. For any $t$, we denote the number of point observed in $(t, t+ dt]$ by $N_{t}$ and for a given dataset as $N_{t,obs}$. 
The windows $\Delta_t$ for the binary approach were selected to create probabilities under an HPP. When counting the number of points, we consider somewhat longer intervals than used for the binary comparison, such that the expected number of points is at least one. 

An important point to make here is that examining the predictive distribution of the number of points in $(t, t+ dt]$ is only appropriate for self-exciting model comparison.  That is, we will compare the predictive distribution  for $N_{t}$ with $N_{t,obs}$.  Self-inhibiting models encourage $N_{t}=0$, we expect decreased intensity (expected number of points) following events.  Thus, in this case, comparing the predictive distribution for $N_t$ to $N_{t,obs}$ is not more informative than the binary comparison discussed above. This was confirmed in our empirical studies for self-inhibiting models.

In this regard, we do model comparison using the posterior predictive distribution for the number of events rather than the posterior mean expected number of events. The latter would encourage the use of, e.g., a predictive mean square error criterion.  The former requires the simulation of posterior predictive realizations and would suggest the use of the ranked probability score (RPS) criterion.
We prefer to employ the RPS \citep{gneiting2007} since it makes comparison with $N_{t,obs}$ using the entire predictive distribution for $N_{t}$ rather than comparison with a point estimate of $E(N_{t})$.

Rather than carrying out the computationally expensive Ogata simulation approach (Algorithm \ref{alg:ogatasim}) many times for each interval $(t, t + dt]$, we use the approximation \eqref{eq:binapprox} as the mean and simulate from a Poisson distribution to generate the posterior predictive distribution $F_{t}$ for $N_t$. Then, we compute RPS$(F_t,N_{t,obs}) = \sum_x (F_t(x) -  \bone(x \geq N_t) )^2  =  \bE|N_t - N_{t,obs} | - \frac{1}{2}\bE | N_t - N'_t |$. The last expression gives expectations that are immediately amenable to Monte Carlo integration using the posterior predictive samples of $N_{t}$. Because we utilize MCMC to fit our Bayesian model, we can obtain such posterior predictive samples.  The resulting Monte Carlo approximation of RPS becomes \citep{kruger2016},
\begin{equation}
\text{RPS}(\hat{F}_t,N_t) = \frac{1}{m} \sum_{j=1}^m |N_{t,j} - N_{t,obs}| - \frac{1}{2m^2} \sum_{j=1}^m \sum_{k=1}^m | N_{t,j} - N_{t,k}| ,
\label{eq:RPS}
\end{equation}
where $m$ is the number of MCMC samples used. Each $t$ has its own $N_{t, obs}$ and predictive distribution for $N_{t}$.  So, we average over the values in \eqref{eq:RPS} to obtain a single model comparison metric. The RPS can be computed in closed form if a single estimate, say the posterior mean, of $\lambda^*(t + dt/2) \, dt$ is used \citep{jordan2017}; however, this approach ignores the uncertainty in estimating $\lambda^*(t + dt/2)$.

\section{Simulation Studies}\label{sec:sim}

\subsection{Identifying Excitation and Inhibition}\label{sec:hpp_sim}

In this section, we present two simulation studies. In the first study, we examine 10 combinations of the parameters $\alpha$ and $\mu$ of a Hawkes process with an exponential triggering function. For each parameter configuration, we simulate 1000 point patterns from that model on $(0,100]$ using Algorithm \ref{alg:ogatasim} in Appendix \ref{app:simulation} and fit a homogeneous Poisson process and an evolutionary point process model to the simulated data. In the second study, we essentially follow the same procedure except we simulate datasets from inhibitory models (i.e., models with $\alpha <0$) using the power link with $\eta =1$. For both self-exciting and inhibitory models, we fix the decay rate to $\beta = 1$. In the studies, we compare the homogeneous Poisson process to an evolutionary point process using DIC and PMR, and we also use RPS for Hawkes processes.

For the Hawkes process simulation, we use combinations of $\mu \in \{0.5,1.0 \}$ and $\alpha \in \{0.01, 0.03, 0.05, 0.07, 0.09 \}$ to examine the degree of baseline activity and excitation needed for an evolutionary model to outperform an HPP. Note that the excitation levels are very low. The results from the simulation study are given in Table \ref{tab:sim_excite}.

We find here that both PMR and RPS reveal significantly better predictive improvement for the self-exciting models relative to the HPP, correctly selecting the Hawkes process models over the HPP almost always across the $1000$ replicate datasets.  PMR improves slightly as $\alpha$ increases.  Neither PMR nor RPS benefit from the roughly doubling of the sample size for $\mu = 1$ vs. $\mu = 0.5$.  DIC seems ineffective.  In fact, it only preferred the Hawkes process more often than the HPP for three parameter combinations: ($\mu = 0.5$, $\alpha = 0.09$); ($\mu = 0.5$, $\alpha = 0.09$); ($\mu = 0.5$, $\alpha = 0.09$).
In all other cases, DIC would have selected the HPP more often than the Hawkes process.

\begin{table}[H]
\centering
\begin{tabular}{l|r|rrr|rrr}
  \hline
 & & \multicolumn{3}{c|}{Averages under Hawkes} & \multicolumn{3}{|c}{Averages under HPP} \\
Parameters & Average $n$ & DIC & PMR & RPS & DIC & PMR & RPS \\
  \hline
$\mu = 0.5$, $\alpha = 0.01$ & 50.571 & 170.546 & 0.304 & 0.488 & 171.110 & 0.356 & 0.519 \\
$\mu = 0.5$, $\alpha = 0.03$ & 51.414 & 171.475 & 0.304 & 0.496 & 172.179 & 0.357 & 0.529 \\
$\mu = 0.5$, $\alpha = 0.05$ & 52.231 & 172.511 & 0.304 & 0.500 & 173.266 & 0.359 & 0.536 \\
$\mu = 0.5$, $\alpha = 0.07$ & 54.188 & 174.585 & 0.301 & 0.508 & 175.569 & 0.362 & 0.552 \\
$\mu = 0.5$, $\alpha = 0.09$ & 55.344 & 175.930 & 0.299 & 0.512 & 176.891 & 0.364 & 0.558 \\
$\mu = 1$, $\alpha = 0.01$ & 100.917 & 200.433 & 0.310 & 0.496 & 200.997 & 0.357 & 0.522 \\
$\mu = 1$, $\alpha = 0.03$ & 103.052 & 200.046 & 0.308 & 0.501 & 200.828 & 0.360 & 0.531 \\
$\mu = 1$, $\alpha = 0.05$ & 105.359 & 199.689 & 0.305 & 0.505 & 200.660 & 0.359 & 0.536 \\
$\mu = 1$, $\alpha = 0.07$ & 107.497 & 199.122 & 0.302 & 0.510 & 200.444 & 0.359 & 0.545 \\
$\mu = 1$, $\alpha = 0.09$ & 110.157 & 198.657 & 0.302 & 0.512 & 199.785 & 0.361 & 0.548 \\
   \hline
\end{tabular}

\caption{Simulation results for mildly self-exciting models. For each model and parameter combination, we present the average number of events $n$, DIC, PMR, and RPS. }\label{tab:sim_excite}

\end{table}

For the inhibitory process simulation, we use combinations of $\mu \in \{2.0, 5.0 \}$ and $\alpha \in \{-0.1, -0.3, -0.5, -0.7, -0.9 \}$ to see which how much deviation from an HPP, in this case toward regularity, is neeeded for an evolutionary model to outperform an HPP. Compared to our Hawkes process simulations, significantly more inhibition was required to effectively distinguish self-inhibiting models from an HPP. The results are presented in Table \ref{tab:sim_inhibit}.

Here, we can effectively select the evolutionary model with both PMR and DIC. In this case, the ability to distinguish models is less dependent on $\mu$, more dependent on $\alpha$. For instance, PMR for the Hawkes model and the HPP are close when $\alpha = -0.1$ but becomes much more consequential as $\alpha$ becomes more negative.  Similar behavior emerges for DIC.


\begin{table}[H]

\centering
%
%
%
%
%
%
\begin{tabular}{lr|rr|rr}
  \hline
 & & \multicolumn{2}{c|}{Averages under Hawkes} & \multicolumn{2}{|c}{Averages under HPP} \\
Parameters & Average $n$ & DIC & PMR & DIC & PMR  \\
  \hline
 $\mu = 2$, $\alpha = - 0.1$  & 181.795 & 138.706 & 0.416 & 147.435 & 0.422 \\
 $\mu = 2$, $\alpha = - 0.3$  & 154.053 & 159.846 & 0.401 & 176.304 & 0.425 \\
 $\mu = 2$, $\alpha = - 0.5$  & 133.997 & 174.070 & 0.372 & 191.070 & 0.427 \\
 $\mu = 2$, $\alpha = - 0.7$  & 118.254 & 179.837 & 0.340 & 198.500 & 0.433 \\
 $\mu = 2$, $\alpha = - 0.9$  & 106.049 & 175.626 & 0.304 & 201.341 & 0.433 \\
 $\mu = 5$, $\alpha = - 0.1$ & 454.586 & -477.183 & 0.415 & -466.340 & 0.419 \\
 $\mu = 5$, $\alpha = - 0.3$ & 386.027 & -284.721 & 0.407 & -269.405 & 0.421 \\
 $\mu = 5$, $\alpha = - 0.5$ & 334.058 & -154.969 & 0.391 & -136.152 & 0.423 \\
 $\mu = 5$, $\alpha = - 0.7$& 295.549 & -67.016 & 0.373 & -47.768 & 0.425 \\
 $\mu = 5$, $\alpha = - 0.9$ & 263.596 & -4.731 & 0.355 & 17.902 & 0.426 \\
   \hline
\end{tabular}
\caption{Simulation results for self-regulating models. For each model and parameter combination, we present the average number of events $n$, DIC, and PMR.}\label{tab:sim_inhibit}
\end{table}

Altogether, we can effectively select the generative evolutionary model that shows excitation and inhibition over the HPP model, even when excitation or inhibition is relatively weak.  PMR emerges as the most effective criterion. 

\subsection{Identifying Link Functions}\label{sec:link_sim}

Here, we provide two different simulation studies. The only difference between the studies is the link function $h(\cdot)$ of the generative model.  Our goal is to explore whether we can effectively distinguish link functions from a simulated point pattern. To do this, we simulate 1,000 point patterns on $(0,100]$ from each generative model using Algorithm \ref{alg:ogatasim} in Appendix \ref{app:simulation}. We then fit each of the simulated datasets with four different models that differ in terms of the link function $h(\cdot)$ used.

Specifically, our generative models use the power link, one with $\eta = 0.5$ and the other with $\eta = 1$. We then fit each simulated point pattern with models using the power link with $\eta = 0.5$ and $\eta = 1$, the soft-plus function, and $\log10$ soft-plus link. For the power link with $\eta = 0.5$, we use $\mu = 3$, $\alpha = 4$, and $\beta= 1$. For the power link with $\eta = 1$, we use $\mu = 0.5$, $\alpha = 0.9$, and $\beta = 1$. These models all exhibit significant excitation regardless of the link function. As before, we use DIC, PMR, and to compare models. The results are presented in Table \ref{tab:sim_link}.  

We find that we cannot well distinguish link functions from a simulated link function using DIC, PMR, or RPS. When simulating from the model using the power link with $\eta = 1/2$, there is almost no separation between any of the models, regardless of the criterion considered. When simulating from the model using the power link with $\eta = 1$, there is almost no separation between model except for the model using the power link with $\eta = 1/2$. The models using the power link with $\eta = 1$, the soft-plus, and $\log_{10}$ soft-plus were hardly indistinguishable in terms of DIC, PMR, and RPS.

\begin{table}[H]

\centering

\begin{tabular}{ll|rrrrr}
  \hline
   & & \multicolumn{4}{c|}{Average} \\
 Simulation Link & Model Link & n & DIC & PMR & RPS \\
  \hline
\multirow{4}{*}{Power, $\eta = 1/2$} & Power, $\eta = 1/2$& 341.748 & -167.983 & 0.283 & 0.555 \\
   & Power, $\eta = 1$ & --- & -168.040 & 0.282 & 0.548 \\
   & Soft-plus & --- & -167.115 & 0.286 & 0.550 \\
   & $\log_{10}$ Soft-plus & --- & -167.190 & 0.285 & 0.550 \\
      \hline
\multirow{4}{*}{Power, $\eta = 1$}  &Power, $\eta = 1/2$ & 460.745 & -796.111 & 0.166 & 0.739 \\
   & Power, $\eta = 1$ & --- & -823.522 & 0.162 & 0.681 \\
   & Soft-plus & --- & -822.086 & 0.162 & 0.678 \\
   & $\log_{10}$ Soft-plus  & --- & -823.444 & 0.162 & 0.681 \\
   \hline
\end{tabular}

\caption{Simulation results for different generative models, differing by the link function. For each generative model and model fit, we present the average number of events $n$, DIC, PMR, and RPS. Dashes indicate that the average number of events $n$ did not change depending on the model fit.}\label{tab:sim_link}

\end{table}

Our explanation is that over the short windows we look at, the link functions produce very similar $\lambda^{*}$'s, hence very similar $q_t$'s. Also, because the number of parameters is the same for all models considered, the likelihoods of the models are very similar. These results suggest that this discrimination task will not be successful given only a single observed point pattern, as will be the case in practice.


\subsection{Evolutionary Log Gaussian Cox Process Models}\label{sec:evo_gp}

In this simulation study, we simulate from a model where the background intensity is that of a log Gaussian Cox process (LGCP) using the power link with $\eta = 1$, i.e., the Tobit link. That is,
\begin{align*}
\mu(t) &= e^{\mu + w(t)}, \\
w(t) &\sim GP\left(0,\sigma^2 e^{- \phi |t - t' |} \right),
\end{align*}
where $\sigma^2 = 1$, $\phi = 1/20$, $\mu = \log(1.5)$, $\alpha = 0.9$, and $\beta = 20$. We then add quickly-decaying excitation to this slowly changing GP in order to help to discriminate between these two components. Using this model, we simulate a dataset $(0,100]$, yielding 1972 events, in total. Because the models considered here are much more computationally expensive to fit, we present the results from the single simulated dataset.

For model fitting, we use prior distributions with relatively high variance compared to the true parameters values: $\mu \sim \text{Normal}(0,10)$, $\alpha \sim \text{Unif}(-2,2)$, $\beta \sim \text{Gamma}(1,1/24)$, $\sigma^2 \sim \text{Inverse-Gamma}\left( 1,1 \right)$, and $\phi \sim \text{Gamma}(10^{-2},10^{-2})$. Note that $\mu$ is no longer constrained to be positive because it is an argument within a natural exponent.

The goal here is to determine whether DIC, PMR, and RPS can be used to select the full model compared to models that exclude the GP or the evolutionary component. The results of this model comparison are given in Table \ref{tab:evo_gp}.

\begin{table}[H]
\centering
\begin{tabular}{rrrr}
  \hline
 & DIC & PMR & RPS \\
  \hline
GP + Evo & -12043.250 & 0.047 & 1.171 \\
GP-only & -12146.550 & 0.088 & 1.302 \\
Evo-only & -11972.690 &  0.050 & 1.121 \\
   \hline
\end{tabular}
\caption{Model comparison criteria for the evolutionary GP model, the GP-only model, and the evolutionary model.}\label{tab:evo_gp}
\end{table}

%

The correct model had the lowest RPS and PMR, albeit by a small margin.  However, both criteria consequentially appreciated the benefit of including the evolutionary component.  In this regard, DIC valued the inclusion of the evolutionary component as well. Because DIC preferred the incorrect model by a substantial margin, it may be preferable to use our proposed intensity-based model comparison criteria when comparing models from different point process classes.
Taking this a bit further, the fitted GP-only model had very fast decay (i.e., large $\phi$) even though the true value of $\phi$ was small. This is likely because the GP-only model required a larger $\phi$ to capture the excitation present in the model. On the other hand, the evolutionary GP model accurately estimated the evolutionary parameters and the GP parameters (see Table \ref{tab:evo_gp_sum}).

\begin{table}[H]
\centering
\begin{tabular}{lrrrr}
  \hline
 & mean & sd & 5\% & 95\% \\
  \hline
$e^{\mu}$ & 1.436 & 0.447 & 0.828 & 2.106 \\
 $\alpha$ & 0.891 & 0.026 & 0.850 & 0.934 \\
  $\beta$ & 19.319 & 1.231 & 17.397 & 21.436 \\
$\sigma^2$ & 1.256 & 0.798 & 0.536 & 2.354 \\
$\phi$ & 0.085 & 0.066 & 0.0199 & 0.219 \\
   \hline
\end{tabular}
\caption{Posterior summaries (mean, standard deviation (sd), as well as the 5th and 95th percentile) for the evolutionary LGCP model.}\label{tab:evo_gp_sum}
\end{table}

\section{Analysis of 2018 Violent Crimes in Chicago}\label{sec:data}

\begin{figure}[H]
\begin{center}
\includegraphics[width=0.45\textwidth]{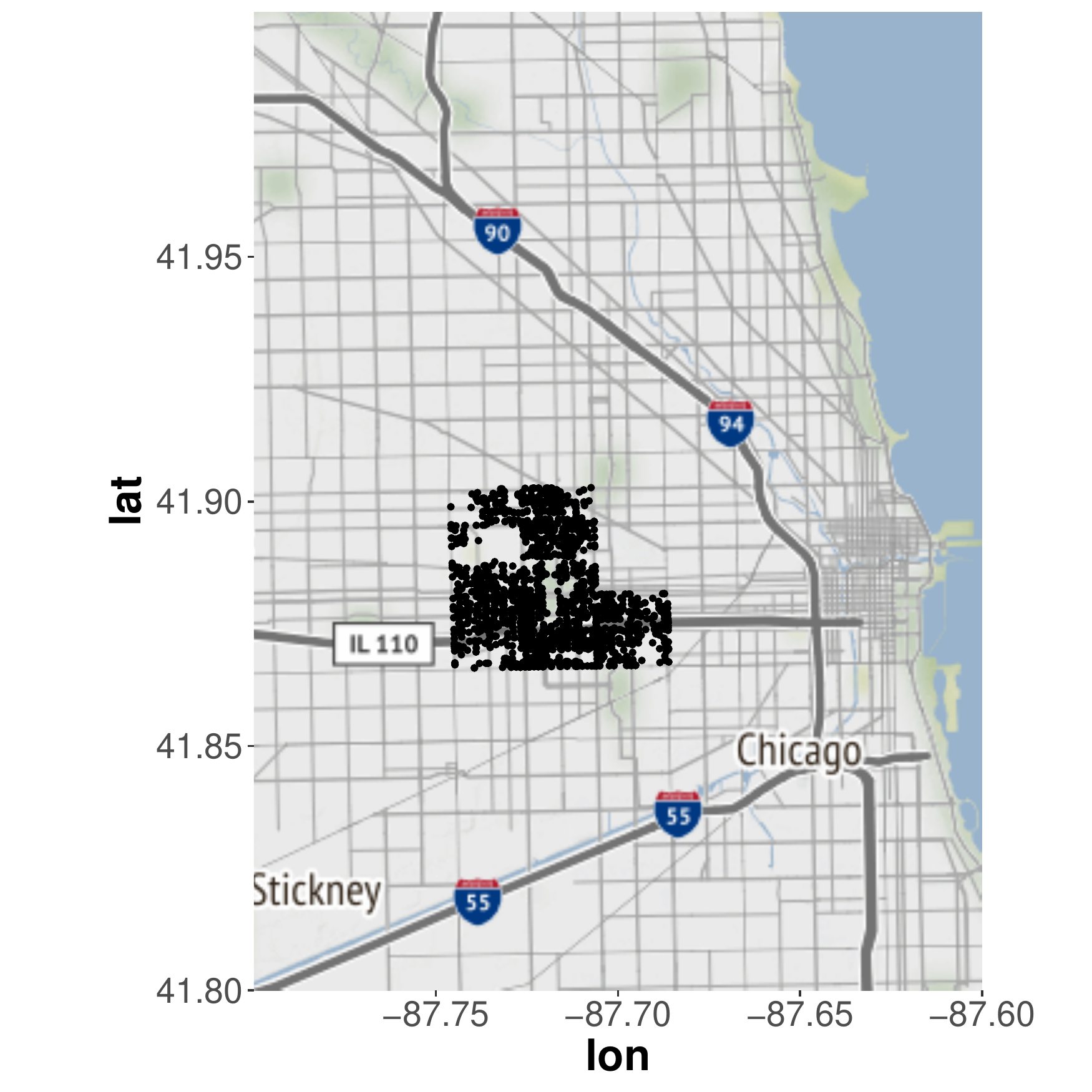}
\end{center}
\caption{Map of violent crime events in police district 11.}\label{fig:crime_locs}
\end{figure}

In this data analysis, we consider a subset of violent crimes - homicides, assaults, and robberies - in Police District 11 of Chicago for the year 2018. These data are publicly available at \url{https://data.cityofchicago.org/Public-Safety/Crimes-2001-to-present}.
Event times are reported to the nearest minute. In total, there are $n = 2448$ events in this dataset, including 41 observations that have the same time as at least one other event \footnote{To avoid ties, we jitter the event times by up to thirty seconds; that is, our data are taken to be a number uniformly distributed between thirty seconds before and thirty seconds after the reported minute.}.   For interpretability of model parameters, we express our data as event times in hours from the beginning of the year. To be clear, we are not rounding event times to the nearest hour; each event keeps date-time information up to the second to which it was jittered. For example, an event recorded at 03:09:00 AM on January 3, 2018, occurs at hour 51.15051. Thus, our data belong to $(0,8760)$.
Although our model is not spatial, we plot the locations of the violent crime events in police district 11 in Figure \ref{fig:crime_locs} to show the small geographic region that District 11 covers. Our primary goal is to assess whether violent crime is self-exciting or, perhaps, self-regulating in this small geographic area.

\begin{figure}[H]
\begin{center}
\includegraphics[width=0.45\textwidth]{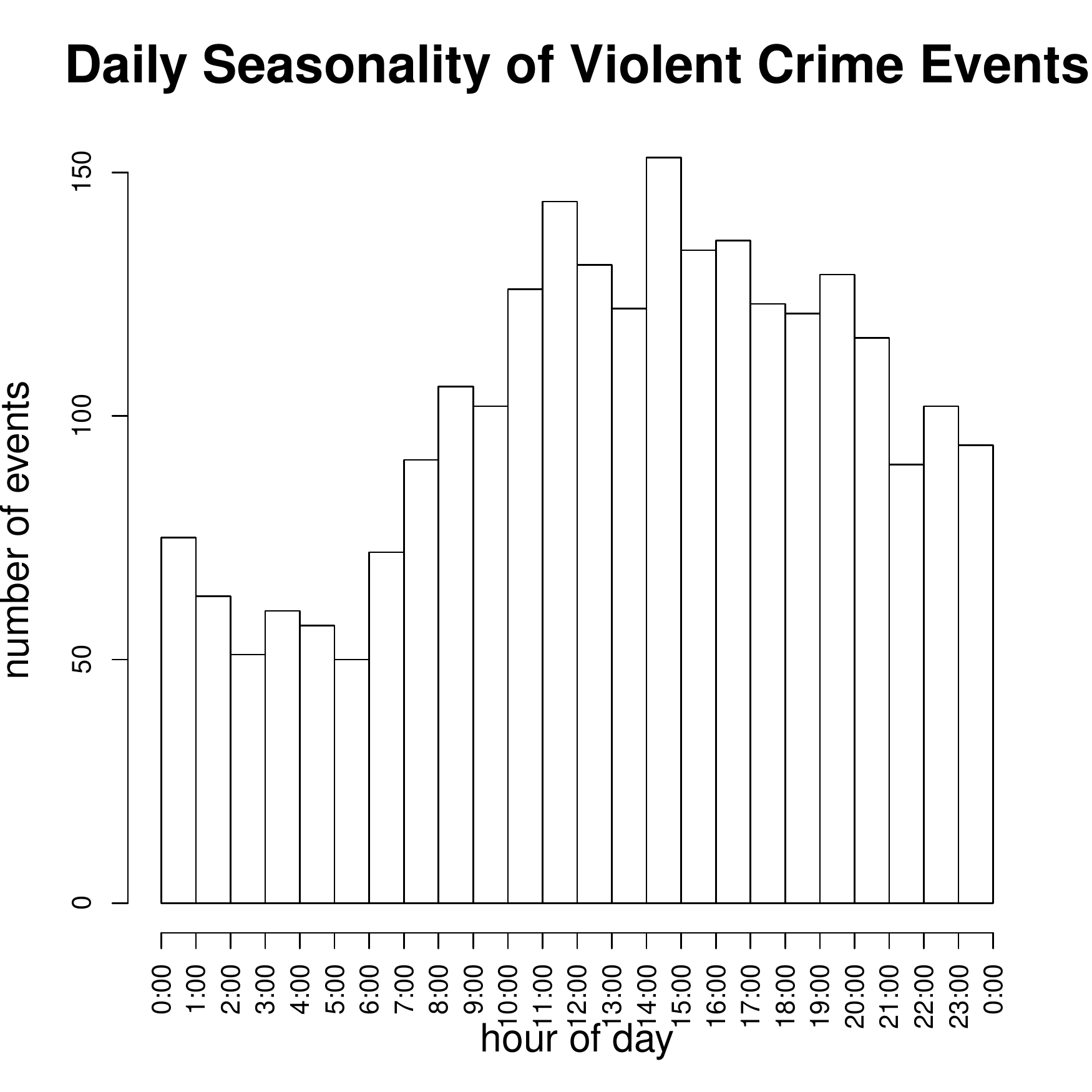}
\includegraphics[width=0.45\textwidth]{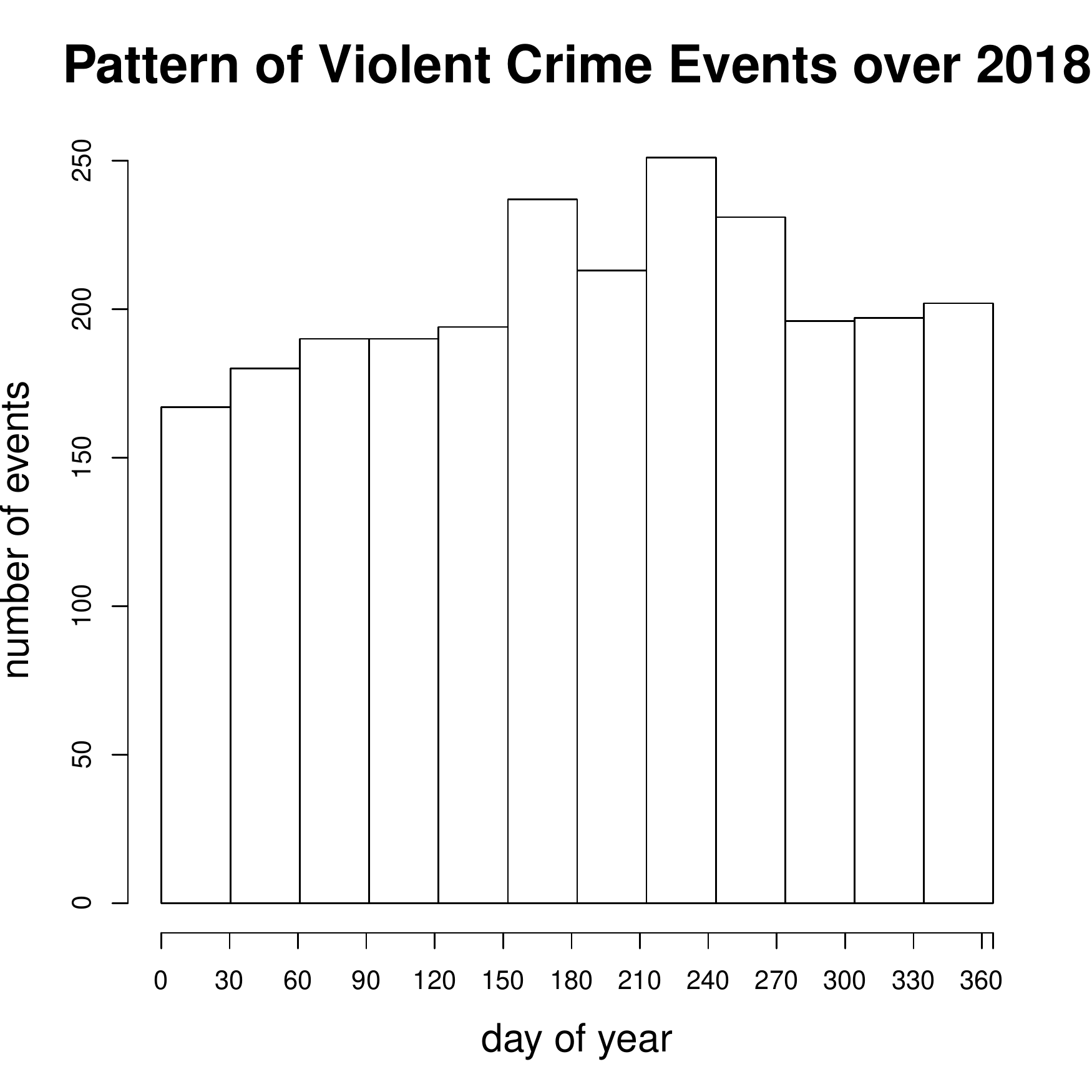}
\end{center}
\caption{Patterns in the Chicago violent crime data (Left) as a function of hour of the day and (Right) over the year. }\label{fig:crime_patterns}
\end{figure}

The data show significant variability as a function of hour of the day and over the year. We illustrate these trends in Figure \ref{fig:crime_patterns}. Specifically, violent crimes happen most often between 10 A.M. and 4 P.M. and are least common 1:00 A.M. and 6 A.M. We also observe higher crime rates in May through September. Unsurprisingly, we see significant deviation from an HPP using a Kolmogorov-Smirnov test ($D = 0.99982, \text{p-value} < 2.2e-16$). However, it is not clear whether the deviation from an HPP can be attributed to excitation/inhibition or daily/annual patterns in the crime data. To account for the clear daily seasonality in event intensity, as well as intensity changes over the year, we argue for modeling the background intensity $\mu(t)$ using seasonal functions and a slowly varying Gaussian process as was used in Section \ref{sec:evo_gp}.

We pose an evolutionary point process model for these data using \eqref{eq:general_univariate}, where we adopt the exponential triggering function for $g(\cdot)$ and the power link with $\eta = 1$ for $h(\cdot)$. We use the following form for the background function:
\begin{align*}
\log(\mu(t)) &= \mu + \gamma_{1} \sin\left(\frac{\pi t}{12} \right) + \gamma)_{2} \cos\left(\frac{\pi t}{12} \right) + w(t) \\
w(t) &\sim GP\left(0,\sigma^2 e^{- \phi |t - t' |} \right).
\end{align*}
Here, the trigonometric terms are used to account for daily seasonality and the log-GP term addresses annual trends in the intensity of events. 

An alternative specification could allow the GP to account for daily seasonality \citep[see, e.g.,]{shirota2017,white2018a}; however, using the exponential covariance function in one-dimension yields a scalable GP model (See Appendix \ref{app:model_fitting}). Because we want $w(t)$ to capture long-term changes in event intensity, we let $\bw^*$ be an evenly-spaced grid of 100 random effects spread over the year. We set $w(t)$ to be equal to the element of $\bw^*$ to which $w(t)$ is closest in time. Thus, $w(t)$ can only capture relatively coarse changes in the event intensity. In this way, short-term changes in intensity are left to be explained by the evolutionary component of the model.


To complete our model specification, we supply prior distributions for all model parameters. Because there is roughly one event for every five or six hours and because the mean parameter $\mu$ is not constrained to be positive, we let $\mu \sim \text{Normal}(0,10)$. Because we allow that crime could be excitatory or inhibitory to the occurrence of future events but do not wish to ``push'' the model in either direction, we let $\alpha \sim \text{Unif}(-2,2)$. We assume that the effect of crime on future events would likely be limited to at most a few days so we let $\beta \sim \text{Gamma}(1,1/24)$. For daily seasonality parameters, we assume that $\gamma_k \sim \text{Normal}\left(0,10 \right)$. In order that the GP captures long-term changes in event intensity, we assume that $\phi \sim \text{Gamma}(1, 50)$. Lastly, we assume $\sigma^{2} \sim \text{Inverse-Gamma}\left( 1,1 \right)$ because this prior distribution is diffuse and gives a closed-form posterior conditional distribution.  

We fit the model using 30,000 MCMC iterations, where we discard the first 10,000 iterations as a burn-in period and use the remaining 20,000 samples for posterior inference. We provide posterior summaries - posterior mean, standard deviation, and 90\% credible interval - for $\mu$, $\alpha$, $\beta$, $\gamma_1$, $\gamma_2$, $\sigma^2$, and $\phi$ in Table \ref{tab:post_sum}. The values of $\alpha$ are small but significantly positive, indicating that there is significant self-excitation present in the data. The relatively large values of $\beta$ suggest that the excitation is short-lived. To give more intuition into how these parameters impact the estimated intensity, we provide some visualizations.

\begin{table}[H]
\centering
\begin{tabular}{lrrrr}
  \hline
 & mean & sd & CI\_0.05 & CI\_0.95 \\
  \hline
$\mu$ & -1.3361 & 0.0212 & -1.3710 & -1.3014 \\
  $\alpha$ & 0.0081 & 0.0022 & 0.0050 & 0.0123 \\
  $\beta$ & 77.3800 & 14.6659 & 49.8852 & 97.8120 \\
  $\gamma_1$ & -0.3081 & 0.0300 & -0.3567 & -0.2589 \\
  $\gamma_2$ & -0.2697 & 0.0286 & -0.3173 & -0.2215 \\
  $\sigma^2$ & 0.8771 & 0.7584 & 0.2600 & 2.1320 \\
  $\phi$ & 1.51e-5 & 1.64e-5 & 1.78e-6 & 4.52e-5 \\
   \hline
\end{tabular}
\caption{Posterior summaries (mean, standard deviation (sd), as well as the 5th and 95th percentile) for the evolutionary LGCP model posed for the Chicago violent crime data.}\label{tab:post_sum}
\end{table}

\begin{figure}[H]
\begin{center}
\includegraphics[width=0.32\textwidth]{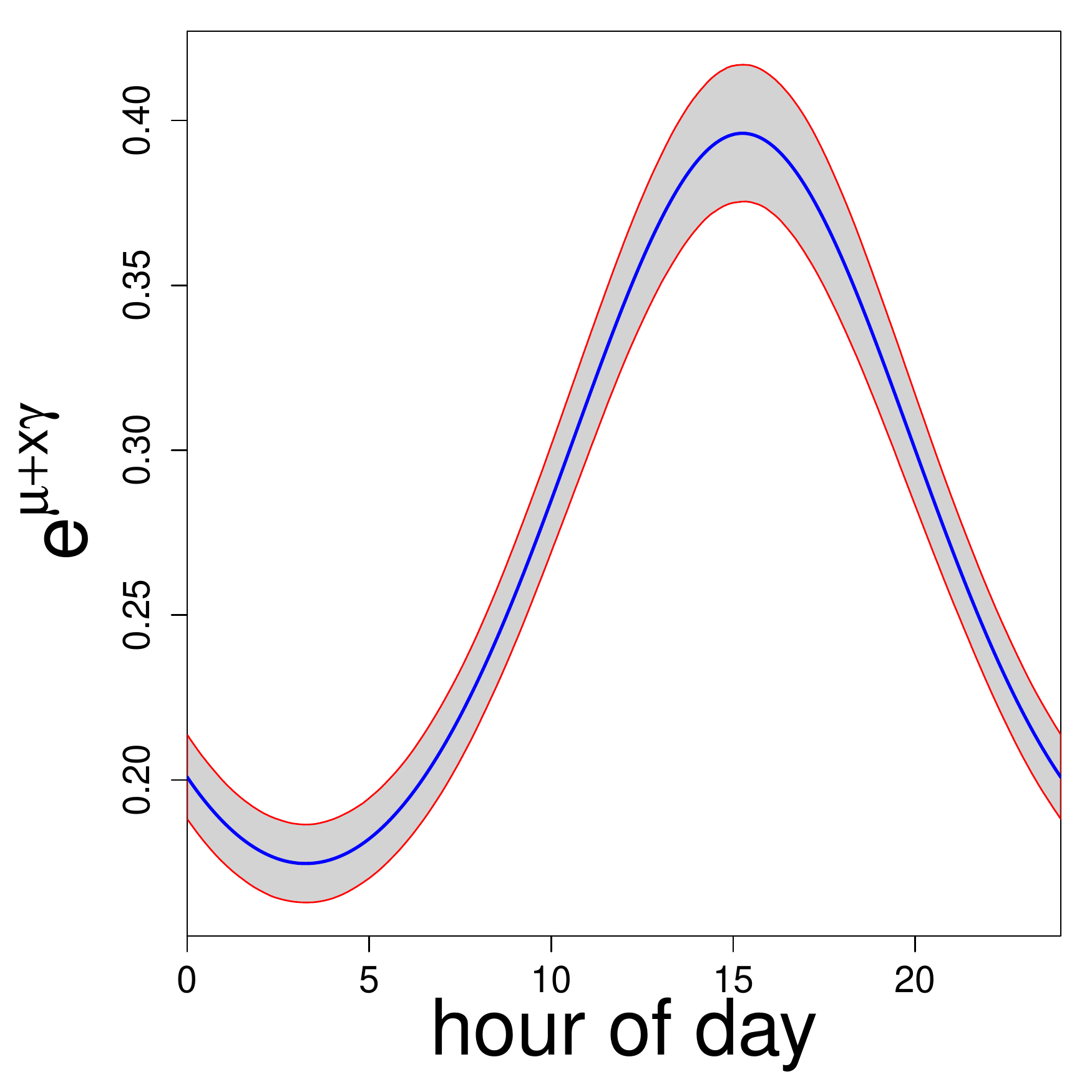}
\includegraphics[width=0.32\textwidth]{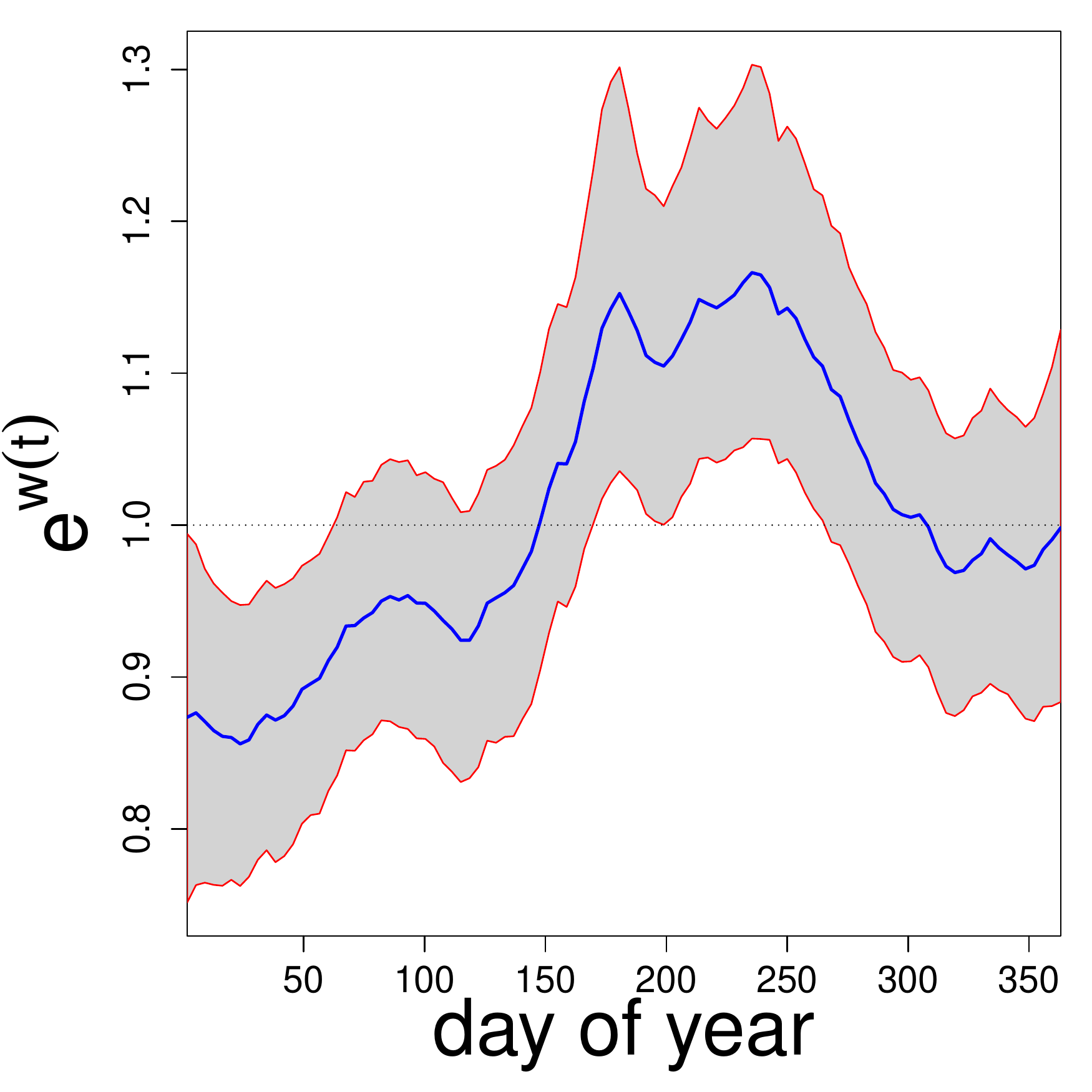}
\includegraphics[width=0.32\textwidth]{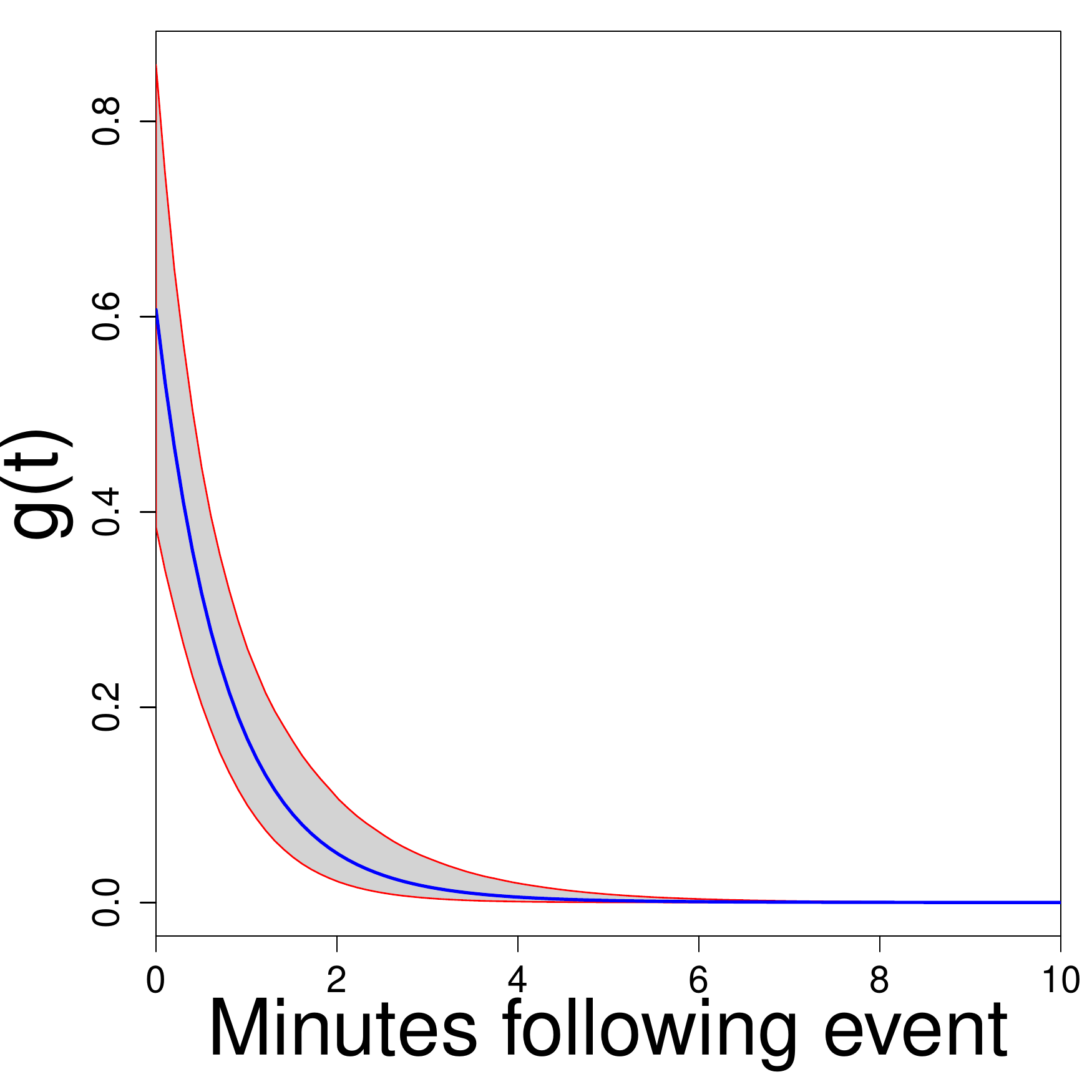}

\end{center}
\caption{(Left) The estimated background intensity, excluding the log GP term, $\exp \left[\mu + \sin\left(\frac{\pi t}{12} \right) \gamma_1 + \cos\left(\frac{\pi t}{12} \right) \right] $. (Center) The estimate log-GP term $e^{w(t)}$ used to capture variation over the year. Here the dotted line shows $e^{w(t)} = 1$  (Right) The estimated evolutionary component of our model, following a single event. This illustrates the pattern of self-excitation estimated by the model. All intensities represent the expected number of events per hour.}\label{fig:fit_curves}
\end{figure}

In Figure \ref{fig:fit_curves}, we provide plots that show the estimated seasonal component of our background intensity, the effect of the log-GP term, and the effect of the evolutionary component of our model following a single event. We first note that the daily seasonality captured by the model is similar to the pattern plotted in Figure \ref{fig:crime_patterns}, suggesting that our model is effectively capturing these daily patterns. In this figure, we estimate that violent crime events occur twice as often in the afternoon compared to the early morning hours. The log-GP also effectively captures the changes over the year that we noted in Figure \ref{fig:crime_patterns}.

There are two periods during the year when the intensity of crime events significantly deviates from the overall pattern. In January and February, we estimate the rate of violent crimes is 10\% less than the average rate of the year. June, July, and August show violent crime rates approximately 15\% above the average rate of the year. This is consistent with the expectation of increased crime rates in warmer weather.
Lastly, we see that the estimated excitation lasts for only a few minutes. With such a short period, this self-excitation could be explained by paired violent events (e.g., fights that lead to multiple assault charges or multiple homicides from gang violence). This explanation is similar to that in \cite{mohler2011} who also model crime with self-exciting point process models. Even though the effect of $\alpha$ is significant in the model, this excitation is very mild.  Given our posterior mean estimate of $\alpha$, we expect a single event to incite 1/100th of an event.

\section{Conclusion}\label{sec:conc}

We have explored model specification and model comparison for
evolutionary point process models within a Bayesian framework. Through
specification of link functions and trigger functions we can
provide flexible models that exhibit excitation as well as inhibition.
Through simulation, we have shown that we can distinguish a
Poisson process specification from either of these behaviors.  However, we can not distinguish link functions within a particular behavior. We have
also enriched the Poisson process to a log Gaussian Cox process and the
evolutionary processes as well, again to attempt to distinguish these
behaviors.   To compare models we have supplied out of sample prediction tools
that enable investigation of short term departure after an event from conditionally independent events arising under a Poisson process.  We have also provided an illustration using a crime dataset.

There is a variety of future work available here.  For example, we could
attempt the use of a spike and slab prior for model comparison, e.g.,
putting a prior on the point mass associated with the Poisson process
and then learning from the posterior about the change in preference for
the Poisson process. See \cite{linderman2014} for some discussion with application to the Hawkes process.  We could also introduce marks.  Such marks could
arise discretely which will lead to evolutionary generalizations of multivariate Hawkes processes \citep[see][for some discussion on this class of models]{chen2017,farajtabar2017}. Alternatively, if we have continuous marks, e.g., spatial
locations, then we could consider evolutionary generalizations of spatio-temporal Hawkes processes \citep[see][for a comprehensive review]{reinhart2017}.

\appendix

\section{Likelihood Derivation}\label{app:likelihood}

We do this in terms of a conditional intensity, but it is trivially applicable to non-evolutionary models. For a point pattern $\mathcal{T}$ over $(0,T]$, we write the joint location density as $f(\mathcal{T}) = f(t_1,...,t_n)$. To define this, we use $f^*(t_i | \mathcal{H}(t_{i})$, the condition density of the $i^\text{th}$ arrival time given previous events, and $F^*(t | \mathcal{H}(t))$, the conditional CDF for any time $t$,

\begin{equation}
F^*(t | \mathcal{H}(t)) = \int^{t}_{t_{k}} f^*(u | \mathcal{H}(t)) du,
\end{equation}

where $t_k$ is the point observed previous to $t$. As we did for intensity, we refer to $f^*(\cdot | \mathcal{H}(t_{i}))$ and $F^*(\cdot | \mathcal{H}(t_{i}))$ as $f^*(\cdot)$ and $F^*(\cdot)$, respectively.

Together, $f^*(\cdot)$ and $F^*(\cdot)$ define the hazard function (the conditional probability of an arrival given the history),

\begin{equation}
\frac{f^*(t)}{1 - F^*(t)},
\end{equation}

which we show is equivalent to the intensity function. We can rewrite the hazard function as
\begin{align*}
\frac{f^*(t)}{1 - F^*(t)}&= \lim_{dt \downarrow 0}  \frac{\Pr(N( (t,t + dt]) > 0 | \mathcal{H}(t) ) /dt }{dt\Pr(N( (t_k,t]) = 0 | \mathcal{H}(t) )} \\
&= \lim_{dt \downarrow 0} \frac{\Pr(N( (t,t + dt]) > 0 ,N( (t_k,t]) = 0   | \mathcal{H}(t) )/dt }{\Pr(N( (t_k,t]) = 0 | \mathcal{H}(t) )} \\
&= \lim_{dt \downarrow 0} \frac{ \Pr(N( (t,t + dt]) > 0 |N( (t_k,t]) = 0  , \mathcal{H}(t))}{dt} \\
&= \lim_{dt \downarrow 0}  \frac{ \bE(N( (t,t + dt]) > 0 | \mathcal{H}(t))}{dt} \\
&=\lambda^*(t).
\end{align*}

Taking this equality and integrating both sides from the point $t_k$ to $t$, where $t_k$ is the point directly previous to $t$,

\begin{align}
\lambda^*(t) &= \frac{f^*(t)}{1 - F^*(t)} = \frac{ \frac{d}{dt} F^*(t)}{1 - F^*(t)} = -\frac{d}{dt} \log\left(1 - F^*(t) \right)  \\
\implies  \int_{t_k}^t \lambda^*(t) dt &= \log\left(1 - F^*(t) \right) - \log\left(1 - F^*(t_k) \right) \nonumber \\
\implies  \int_{t_k}^t \lambda^*(t) dt &= \log\left(1 - F^*(t) \right). \nonumber
\end{align}

By our assumption that the point pattern is simple, no points can appear exactly simultaneously. Thus, $F^*(t_k) = 0$. In words, observing an additional point given that we have already observed an event at $t_k$ has zero probability. Thus, the last implication holds. We can rearrange these to get

\begin{align}
F^*(t) &= 1 - \exp\left( - \int^t_{t_k} \lambda^*(u) du \right) \\
f^*(t) &= \lambda^*(t) \exp\left( - \int^t_{t_k} \lambda^*(u) du \right)  \nonumber.
\end{align}

These are then used to specify the joint likelihood of the observed point pattern.

We define $t_0 = 0$, and because $\mathcal{T} \in (0,t_n] \subset (0,T]$, we must also include the probability of not observing an event in $(t_n,T]$, $(1 -F^*(T))$. We can now decompose the joint density of the point pattern $f(t_1,...,t_n)$ conditionally as

\begin{align}
f(\mathcal{T}) = f(t_1,...,t_n) &= (1 -F^*(T)) \prod^n_{i=1} f^*(t_i ) , \\
&= \exp\left( - \int^T_{t_n} \lambda^*(u) du \right) \prod^n_{i=1} \lambda^*(t_i) \exp\left( - \int^{t_i}_{t_{i-1}} \lambda^*(u) du \right), \nonumber \\
&= \exp\left( - \int^T_{0} \lambda^*(u) du \right) \prod^n_{i=1} \lambda^*(t_i). \nonumber
\end{align}

\section{Model Fitting}\label{app:model_fitting}

Here, we present the Markov chain Monte Carlo sampler for the evolutionary LGCP model presented in Section \ref{sec:data}. To fit simpler specifications where $\mu(t) = \mu$, one only needs to exclude updates to the GP components of the model. As discussed, our model fitting approach resembles the conditional intensity approach in \cite{rasmussen2013} because there is no cluster representation because our models may include inhibition that violates the Poisson additivity required for the cluster representation. To keep our models scalable, we have used the exponential triggering function which is amenable to recursive calculations \citep{ogata1978} shown below.  Here, the log-likelihood of this model is
\begin{equation*}
-\Lambda^*(T) + \sum_{i = 1}^n \log \lambda^*(t_i),
\end{equation*}
where
\begin{align*}
\Lambda^*(T) &= \int^T_0 \lambda^*(s)  ds \\
\lambda^*(s) &= h\left(  \exp\left[\mu + \sin\left(\frac{\pi s}{12} \right) \gamma_1 + \cos\left(\frac{\pi s}{12} \right) \gamma_2  + w(s)\right] + \alpha \beta \sum_{t_i < s} e^{ - \beta (s - t_i) }\right).
\end{align*}
When $s = t_i$, $\alpha \beta \sum_{t_j < t_i} e^{ - \beta (t_i - t_j) } = \alpha \beta A_i,$ where $A_i = e^{-\beta (t_i - t_{i-1}) }\left(1 + A_{i-1} \right)$ and $A_1 = 0$. This enables us to compute $\sum_{i = 1}^n \log \lambda^*(t_i)$ on $\mathcal{O}(n)$.

The compensator $\Lambda(T)$ cannot be computed in closed-form as in \cite{rasmussen2013} because we cannot simply bring the integral inside the non-linear link (operator) $h(\cdot)$. Here, we compute the integral numerically using the trapezoidal rule $$\Lambda^*(T) \approx \sum^{K}_{k=1 } \frac{\lambda^*(s_{k-1}) + \lambda^*(s_{k})}{2}(s_k - s_{k-1}),$$ where $s_0 = 0$, $s_K = T$, and $s_k$ are evenly spaced between $s_0$ and $s_K$. For this approximation, using the nearest $t_i < s$ we can use $\alpha \beta \sum_{t_j < s} e^{ - \beta (s - t_j) } = \alpha \beta e^{-\beta (s - t_{i}) }\left(1 + A_{i} \right)$ to compute $\lambda^*(s)$ with a computational cost that is linear in $K$. Thus, we choose $K$ to be large, $\approx 10^4$.

 For computationally efficient GP prior distributions, we adopt the exponential covariance function.  This corresponds to a continuous-time AR(1) process \citep[see, e.g.,][]{brockwell2007,white2019_multi}. Hence, the joint distribution of $w(t)$ can be peeled off sequentially, i.e.,
\begin{equation}\label{eq:ar_ou}
[w(t_1),..., w(t_{s}),...]=
[w(t_1)][w(t_2)|w(t_1)]...[w(t_{s})|w(t_{s-1})]...
\end{equation}
where $[w(t_{s})|w(t_{s-1})] = N(e^{-\phi(t_{s} - t_{s-1})}w(t_{s-1}),1 -e^{-2 \phi (t_s - t_{s-1}) })$ with $[w(t_0)] = N(0,1)$. Thus, even though we cannot integrate out $w(t)$, computing the prior distribution is not computationally expensive. To sample from the posterior of each random effect, we use the Metropolis algorithm with a Normal random walk, each with its own candidate variance tuned for acceptance rates between 0.15 and 0.6.

For a proper log-Gaussian Cox process model, we must evaluate the GP at each observed event time and over a grid to compute the compensator, which we do for Section \ref{sec:evo_gp}. However, for Section \ref{sec:data}, we use a 100 random effects, denoted $\bw^*$ spaced evenly across (0,T] so that the GP does not absorb possible short-term excitation in the data. Then, we let $w(t)$ be equal to the element to $\bw^*$ that is closest to $w(t)$ in time.

Again, because of the non-linear link $h(\cdot)$, full conditional distributions cannot be derived in closed form. Therefore, we sample from the posterior distribution of $\mu$, $\alpha$, $\beta$, $\gamma_1$, and $\gamma_2$ using the Metropolis algorithm using a Normal random walk proposal distribution, each with unique candidate variance. We use a burn-in of 10,000 iterations. For the first half of the burn-in period plus a period of 100 iterations (5,100 iterations, in this case), we sample from $\mu$, $\alpha$, $\beta$, $\gamma_1$, and $\gamma_2$ individually and tune the candidate variance so that acceptance rates are between 0.15 and 0.6. After this time when the Markov chain has had time to converge, we change our proposal distribution to a multivariate normal distribution with covariance is calculated using posterior samples collected after the first half of the burn-in period (after the 5,000th iteration). The covariance of the proposal distribution is updated throughout the entire sampler to ensure its convergence \citep{haario1999}. We found that this combination of univariate and then multivariate updates was effective in sampling from the posterior distribution of the model.

\section{Simulation of Evolutionary Point Processes}\label{app:simulation}

Because we rely heavily on simulation studies to illustrate the utility and limitations of our model comparison discussion. We provide a general approach for simulating evolutionary point patterns from \cite{rasmussen2011} based on the thinning methods from \cite{ogata1981} in Algorithm \ref{alg:ogatasim}. This algorithm allows for both excitation and inhibition.

\begin{algorithm}[H]
\footnotesize

\caption{Ogata's Thinning Method for Generating Evolutionary Point Process  \citep{ogata1981}}\label{alg:ogatasim}

 \hspace*{\algorithmicindent} \textbf{Input}:
 \begin{itemize}
  \itemsep0em
   \item conditional intensity function $\lambda^*(\cdot)$
 \item maximum proposal time $l(t)$
 \item termination time $T_\text{end}$
\end{itemize}

 \hspace*{\algorithmicindent} \textbf{Output}:
  \begin{itemize}
  \itemsep0em
   \item A realization of an evolutionary point process
\end{itemize}

\begin{algorithmic}[1]
   \State $\mathcal{T} = \emptyset$; $n \gets 0$; $t \gets 0$
   \While{$t <T_\text{end}$}:
   \State Take $\tilde{\lambda} \geq \sup_{ s \in [t, t + l(t)] } \lambda^*(s) $
   \State Draw $w \sim \text{Exponential}(m(t))$
   \If{$w > l(t)$}:
   \State $t = t + l(t)$
   \Else
   \State Set $t = t + w$
   \State Generate $U \sim \text{Unif}(0,1)$
   \If{$U \tilde{\lambda} \leq \lambda^*(t)$}:
   \State $n \gets n + 1$
   \State $t_n \gets t$
   \State $\mathcal{T} \gets \mathcal{T} \cup \{ t_n\}$
   \EndIf
   \EndIf
   \EndWhile
   \If{$t_n \leq T$}:
   \State \Return $\mathcal{T}$
   \Else:
   \State \Return $\mathcal{T} \setminus \{ t_n \}$
      \EndIf
\end{algorithmic}

\end{algorithm}

\bibliographystyle{apalike}
\bibliography{refs}

\end{document}